\documentclass{LMCS}

\def\dOi{10(2:12)2014}
\lmcsheading%
{\dOi}
{1--10}
{}
{}
{Nov.~20, 2013}
{Jun.~19, 2014}
{}


\ACMCCS{[{\bf Theory of computation}]: Computational complexity and
  cryptography---Complexity classes} 

\usepackage{algorithmic,hyperref}

\begin{document}

\title[Unsolvability Cores in Classification Problems]
      {Unsolvability Cores in Classification Problems}

\author[U.~Brandt]{Ulrike Brandt}
\author[H.~K.-G.~Walter]{Hermann K.-G. Walter}
\address{Fachbereich
  Informatik, Technische Universit\"{a}t Darmstadt, Germany}
\email{brandt@dekanat.informatik.tu-darmstadt.de,  
  walter@informatik.tu-darmstadt.de} 

%
%
\begin{abstract}
Classification problems have been introduced by M.~Ziegler as a
generalization of promise problems. In this paper we are concerned
with solvability and unsolvability questions with respect to a given
set or language family, especially with cores of unsolvability. We
generalize the results about unsolvability cores in promise problems
to classification problems. Our main results are a characterization of
unsolvability cores via cohesiveness and existence theorems for such
cores in unsolvable classification problems. In contrast to promise
problems we have to strengthen the conditions to assert the existence
of such cores. In general unsolvable classification problems with more
than two components exist, which possess no cores, even if the set
family under consideration satisfies the assumptions which are
necessary to prove the existence of cores in unsolvable promise
problems. But, if one of the components is fixed we can use the
results on unsolvability cores in promise problems, to assert the
existence of such cores in general. In this case we speak of
conditional classification problems and conditional cores. The
existence of conditional cores can be related to complexity
cores. Using this connection we can prove for language families, that
conditional cores with recursive components exist, provided that this
family admits an uniform solution for the word problem.
\end{abstract}
\maketitle
\section*{Introduction}
The concept of \textit{classsification problems} was introduced by M. Ziegler ([1]) as a generalization of \textit{promise problems} due to S. Even ([5]). Promise problems are a generalization of \textit{decision problems}. A classification problem is a vector $ \textbf{A}=(A_1,\dots, A_k)$ where the $A_i$ are pairwise disjoint infinite subsets of a given basic set $S$. For a set family $\mathcal{F} \subseteq \textbf{2}^S$ such a classification problem is  $ \mathcal{F}$-\textit{solvable}, if a vector $\textbf{Q}=(Q_1,\dots, Q_k)$ exists with $A_i \subseteq Q_i$, $Q_i \in \mathcal{F}, Q_i \cap Q_j = \emptyset$ for $ 1\leq i \neq j \leq k $ and $ Q_1 \cup \dots \cup Q_k = S.$ If $ k=2 $ we are faced with promise problems. In applications $ S= X^*$ where $X$ is a finite nonempty alphabet and $ \mathcal{F} = \mathcal{L}$ a language family and/or a complexity class. From an algorithmic point of view solutions of classification problems can be used to obtain constant size \textit{advices}. In this case advices indicate the inputs to belong to certain subsets (c.f. [1] for further details). We extend the results about unsolvability cores in promise problems ([4]) to unsolvability cores in classification problems. Again cohesiveness is the characterizing indicator. For unsolvable promise problems we can find in general unsolvability cores, if the set family is closed under union, intersection and finite variation. But for unsolvable classification problems with $k > 2$ the existence of unsolvability cores needs further conditions. We show, that we can assert the existence of unsolvability cores for $ k > 2$ under the same assumption as needed for promise problems, if we fix one of the components. In this approach the fixed component is called the \textit{condition} for the classification problem. The results are proven under assumptions which involve closure properties of $\mathcal{F}$ against some or all boolean operations \textit{union, intersection} and \textit{complementation}. Moreover, we can relate unsolvability cores for conditional classification problems to so called \textit{proper hard cores} introduced by R. Book and D.-Z. Du in a general form ([3]) and first defined by N. Lynch ([6]) for complexity classes. Using results and proof techniques from [3] we can apply our results to language families and complexity classes.  Especially, we are able to construct unsolvability cores where the components are recursive. To do this, the language family or complexity class under consideration must allow an enumeration where the word problem has a uniform solution. We assume the reader to be familiar with the theory of recursive functions, languages and complexity (cf.[2],[7]).

\section{Set and Language Families, Basic Notations}
In the following an infinite \textit{basic} set $S$ is given. We assume that the elements of set families $\mathcal{F}$ are subsets of S. Moreover, sets $A, A', B, B', C, \cdots, Q, \cdots$ are always subsets of $S$ and \textit{singletons} $\{s\}$ are identified with $s$. 
We mainly deal with denumerable set families $\mathcal{F}$; i.e. a function $\textbf{e}_\mathcal{F}:\ \mathbb{N}_0 \rightarrow \textbf{2}^S$ with $\textbf{e}_\mathcal{F}(\mathbb{N}_0)=\mathcal{F}$ exists (\textit{enumeration} of $\mathcal{F}$). Consider the \textit{boolean} operations $A \cup B$  \textit{union}, $A \cap B$ \textit{intersection} and $A^\textbf{c} = S $\textbackslash $A$ \textit{complementation} in connection with set families $\mathcal{F}$. These operations can be lifted to binary operations between set families $\mathcal{F}_1$ and $\mathcal{F}_2$ and unary operations for $\mathcal{F}$. Define
\begin{align*}
&\mathcal{F}_1 \oplus \mathcal{F}_2=\{A \cup B | A \in \mathcal{F}_1 \ \text{and} \  B \in \mathcal{F}_2\}, \\
&\mathcal{F}_1 \odot \mathcal{F}_2 = \{A \cap B | A \in \mathcal{F}_1\ \text{and} \  B \in \mathcal{F}_2\}
\end{align*}
and the closure operations
\begin{align*}
&\mathcal{F}^\textbf{u} = \{A_1 \cup \ldots \cup A_n | n \geq 1, A_i \in \mathcal{F}\ for \ 1 \leq i \leq n\}(union),\\
&\mathcal{F}^\textbf{s} = \{A_1 \cap \ldots \cap A_n | n \geq 1, A_i \in \mathcal{F}\ for \ 1 \leq i \leq n\}(intersection),\\
&\mathcal{F}^\textbf{co} = \{A^\textbf{c} | A \in \mathcal{F}\},\ \mathcal{F}^\textbf{cc} = \mathcal{F} \cup \mathcal{F}^\textbf{co} (complementation)  \ \text{and} \\
&\mathcal{F}^\textbf{b} = ((\mathcal{F}^\textbf{cc})^\textbf{s})^\textbf{u} (boolean\ closure).
\end{align*}
We will frequently use $\mathcal{F}^\textbf{dc} = \mathcal{F} \cap \mathcal{F}^\textbf{co}$. 
Note, that $(\mathcal{F}^\textbf{u})^\textbf{s} = (\mathcal{F}^\textbf{s})^\textbf{u} (distributivity),  (\mathcal{F}^\textbf{co})^\textbf{u} = (\mathcal{F}^\textbf{s})^\textbf{co} (deMorgan)$, 
$(\mathcal{F}^\textbf{cc})^\textbf{dc} =\mathcal{F}^\textbf{cc}$ and $(\mathcal{F}^\textbf{co})^\textbf{co} = \mathcal{F}$. Furthermore, $\mathcal{F}= \mathcal{F}^\textbf{cc}$ ( $\mathcal{F}= \mathcal{F}^\textbf{u}$, $\mathcal{F}= \mathcal{F}^\textbf{s}$) if and only if $\mathcal{F}= \mathcal{F}^\textbf{co}$ ($\mathcal{F} \oplus \mathcal{F} \subseteq \mathcal{F}$, $\mathcal{F} \odot \mathcal{F} \subseteq \mathcal{F}$, respectively).

 Let $\textbf{\textit{fin}}(S) = \{A \subseteq S | A $ finite$\}$. Then $\mathcal{F}$ is closed under\textit{ finite variation} if $\mathcal{F} \oplus \textbf{\textit{fin}}(S)\subseteq\mathcal{F}$ and $\mathcal{F} \odot \textbf{\textit{fin}}(S)^\textbf{co}\subseteq\mathcal{F}$. We call $\mathcal{F}$ \textit{nontrivial} if $\emptyset, S\in\mathcal{F}$ and $\mathcal{F}$ is closed under finite variation.
In this case $\textbf{\textit{fin}}(S) \subseteq \mathcal{F}$. Note, that $\textbf{\textit{fin}}(S) = \textbf{\textit{fin}}(S)^\textbf{b}$. Moreover, $\mathcal{F}^\textbf{cc}$, $\mathcal{F}^\textbf{u}$, $\mathcal{F}^\textbf{s}$ and $\mathcal{F}^\textbf{b}$ are nontrivial, if $\mathcal{F}$ is nontrivial. \\
Consider the case $S = X^\ast$, where $X^\ast$ is the \textit{free monoid over}$X$ (a nonempty, finite \textit{alphabet}) with \textit{concatenation} of \textit{words} as monoid operation and \textbf{1} as identity. As usual $L \subseteq X^\ast$ is called a \textit{language} and $\mathcal{L} \subseteq \textbf{2}^{X^\ast}$ a \textit{language family}. 
 For a word $w = x_1 \ldots x_n$ $(x_i \in X$ for $1 \leq i \leq n)$  $|w|=n$ is the \textit{length} of $w$ and $|\textbf{1}| = 0$.
For languages $L_{1}$ and $L_{2}$ the \textit{complex product} is defined by $L_1L_2 = \{w_1w_2 | w_1 \in L_1, w_2 \in L_2\}$. There are various kinds of quotients available, for example the \textit{left quotient} defined by $L_1^{-1}L_2 = \{w |\ \exists w_1 \in L_1$: $w_1w \in L_1 \}$. In this context we are mainly interested in handling \textit{leftmarkers}, i.e. we consider the products $wL$ and the quotients $w^{-1}L$ where $ w\in X^\ast$ and $L$ is a language. With respect to language families $\mathcal{L}$ we get the closure operations  $\mathcal{L}^\textbf{ltr} = \{wL | w \in X^\ast, L \in \mathcal{L}\}$ and 
$\mathcal{L}^\textbf{-ltr} = \{w^{-1}L | w \in X^\ast, L \in \mathcal{L}\}$.
In handling the leftmarkers (for example complementation of a leftmarked language) we use variation by  $\mathcal{L}_{\textbf{reg}}(X)$, the family of \textit{regular} languages (for details see [4]). A language family $\mathcal{L}$ is closed under \textit{regular variation} if $\mathcal{L} \oplus \mathcal{L}_{\textbf{reg}}(X) \subseteq \mathcal{L}$ and $\mathcal{L} \odot \mathcal{L}_{\textbf{reg}}(X) \subseteq\mathcal{L}$.

Looking at (partial) orderings on $X^\ast$ the \textit{lexicographic} ordering is important for our purposes. For $ n \geq 0$ let $[n]_0 = \{0,  \dots , n - 1\}$ and $[n] = \{1, \dots ,n\}$. Given a bijection $\omega:X \rightarrow [b]_0 \ (b=\#(X))$ define  
$w \leq v$ if and only if ($|w| < |v|$ or ($|w| = |v|$ and $(\forall u \in X^\ast, x,y \in X:\ w \in uxX^\ast$ and $\ v \in uyX^\ast \Rightarrow \omega(x) \leq \omega(y)))$. 
This is a well-ordering, hence we can define a successor function $\textbf{\textit{succ}}$ for $w \in X^\ast$ by 
$\textbf{\textit{succ}}(w) = \textbf{\textit{min}}\{v \in X^\ast | w \neq v $ and $ w \leq v\}$
where the minimum is taken with respect to the lexicographic ordering. Then
$\lambda i. \textbf{\textit{lex}}(i) = \textbf{\textit{succ}}^i(\textbf{1})$ defines a bijection $\textbf{\textit{lex}}:\ \mathbb{N}_0 \rightarrow X^\ast$ with inverse $\textbf{\textit{ord}} =\textbf{\textit{lex}}^{-1}$. 

Consider the language families $\mathcal{L}_{\textbf{r.e.}}(X)$ (\textit{recursively enumerable} languages) and  $\mathcal{L}_{\textbf{rec}}(X)$ \\$=\mathcal{L}_{\textbf{r.e.}}(X)^\textbf{dc}$ (\textit{recursive} languages). Let $\textbf{\textit{rec}}_n (n \geq 0)$ be the set of \textit{n-ary recursive} functions. 
Using $0, 1\in \mathbb{N}_0$ as truth values define for a language $L$ the function $ \lambda i.\delta_L(i) = "\textbf{\textit{lex}}(i) \in L"$. Then a language $L$ is recursive if and only if $\delta_L \in \textbf{\textit{rec}}_1\ $. Alternatively, a nonempty language $L$ is recursive if and only if a function $f :\mathbb{N}_0 \rightarrow X^*$ exists such that  $ \lambda i.\textbf{\textit{ord}}(f(i))$ is nondecreasing and recursive. 
Classical language families and complexity classes are always denumerable. Of special interest are families with enumerations which are in a certain sense "effective". For our purpose it is important to assert that these enumerations allow a uniform solution for the word problem. More formular, we define for an enumeration \textbf{e} of a language family $\mathcal{L}$ the function  $\lambda i,j$.$\textbf{\textit{word}}_\textbf{e}(i, j) = "\textbf{\textit{lex}}(j) \in \textbf{e}(i)"$. If $\textbf{\textit{word}}_{\textbf{e}} \in \textbf{\textit{rec}}_2$ then \textbf{e} is called \textit{WP-recursive}. $\mathcal{L}$ is called \textit{WP-recursive}, if a WP-recursive enumeration \textbf{e} of $\mathcal{L}$ exists.  Note, that any WP-recursive $\mathcal{L}$ is a (proper) subfamily of $\mathcal{L}_{\textbf{rec}}(X)$ and every complexity class with reasonable ressource bounds (\textit{time-} and \textit{space-constructability} [2]) is WP-recursive.

\section{Solvability of Classification Problems}
Let $k > 0$. We consider vectors $  \textbf{A}=(A_1,\dots, A_k)$ with $ A_i \subseteq S$ for $ 1\leq i \leq k $. To such an \textbf{A} we associate two functions $ \textbf{\textit{set}}(\textbf{A}) = A_1 \cup\dots\cup A_k$ and $|\textbf{A}| = k$. Moreover, if $  \textbf{B}=(B_1,\dots, B_m)$ with $1 \leq m\leq k$ is another vector, then $ \textbf{B} \leq \textbf{A}$ if and only if an injective  $\sigma:[m] \rightarrow [k] $ exists with $ B_i \subseteq A_{\sigma (i)}$ for $ 1 \leq i \leq m$. \textbf{A} is a \textit{classification problem} if $A_i$ is infinite and $ A_i \cap A_j = \emptyset$ for all  $ 1\leq i \neq j \leq k $. For a given $ \mathcal{F}$ a vector $ \textbf{Q}=(Q_1,\dots, Q_k)$ is an $\mathcal{F}$ \textit{-partition} if $ \textbf{\textit{set}}(\textbf{Q}) = S$, $Q_i \in \mathcal{F}$ and $Q_i \cap Q_j = \emptyset$ for $ 1\leq i \neq j \leq k $.

\begin{defi} A classification problem $\textbf{A}$ is $\mathcal{F}$-\textit{solvable} ($A\in\textbf{\textit{class}}_k(\mathcal{F})$) if and only if an $\mathcal{F}$-partition \textbf{Q} exists with $|\textbf{Q}| = k$ and $ \textbf{A} \leq \textbf{Q}$, where $k = |\textbf{A}|$.
\end{defi}

If $S = \mathbb{N}_0$ then $\mathcal{F}$-solvability of promise problems corresponds to the \textit{separation principle} defined in [7] (exercise 5-33). Our definition of  $\mathcal{F}$-solvability for classification problems is stronger than the definition of $\mathcal{F}$-\textit{separability} given in [1], where a classification problem $\textbf{A}$ is $\mathcal{F}$-\textit{separable}, if there exists a $\textbf{Q}$, which satisfies the conditions of Definition 2.1. except the condition "$ \textbf{\textit{set}}(\textbf{Q}) = S$", which may not necessarily be valid. Note that for such a $\textbf{Q}$, we always obtain $ Q_k \subseteq (Q_1 \cup\dots\cup Q_{k - 1})^\textbf{c}$. Hence, the class of $\mathcal{F}$-solvable classification problems with more than one components is identical with the class of $\mathcal{F}$-separable classification problems, if $\mathcal{F}$ is a boolean algebra. That $\mathcal{F}$-solvability is stronger than $\mathcal{F}$-separability, follows from results in [7]. Consider $\mathcal{L}_{\textbf{r.e.}}(X)$ where X is a one-letter alphabet. Then a promise problem $(A, B)$ consisting of recursively enumerable sets exists, which is not $\mathcal{L}_{\textbf{r.e.}}(X)$-solvable ([7] exercise 5-34). But $(A, B)$ is clearly $\mathcal{L}_{\textbf{r.e.}}(X)$-separable. We also find the interesting result that any promise problem $(A, B)$ with $ A, B \in \mathcal{L}_{\textbf{r.e.}}(X)^\textbf{co}$ is  $\mathcal{L}_{\textbf{r.e.}}(X)^\textbf{co}$-solvable ([7] exercise 5-33). Hence all promise problems, which are $\mathcal{L}_{\textbf{r.e.}}(X)^\textbf{co}$-separable are $\mathcal{L}_{\textbf{r.e.}}(X)^\textbf{co}$-solvable. But $\mathcal{L}_{\textbf{r.e.}}(X)^\textbf{co}$ is not closed under complementation.  

For $k = 1$ we identify $A_1$ with $(A_1)$. If $\mathcal{F}$ is nontrivial then every $A_1$ is $\mathcal{F}$-solvable. If $k > 2$ and $\mathcal{F}$ satisfies appropriate closure properties, then we can reduce the question of solvability of classification problems to solvability of promise problems. Directly from the definition we get

\begin{prop} \text{}
If $\mathcal{F} = \mathcal{F}^\textbf{u}$ then for all classification problems \textbf{A} and \textbf{B} with $\textbf{B} \leq \textbf{A}$  $\textbf{A} \in \textbf{class}_{ \mid \textbf{A} \mid} ( \mathcal{F}) $ implies $ \textbf{B} \in \textbf{class}_{\mid\textbf{B}\mid}(\mathcal{F})$.
\end{prop}
\proof Suppose $\textbf{B} \leq \textbf{A} \leq \textbf{Q}$ where $\textbf{Q}$ is an $\mathcal{F}$-partition. Let $B=(B_1, \dots ,B_m), A=(A_1, \dots ,A_k)$ and $Q=(Q_1, \dots ,Q_k)$. Then we can assume without loss of generality $B_i \subseteq A_i \subseteq Q_i$ for all i. Consider P = $Q_1 \cup \dots \cup Q_k$. Then $P^\textbf{c} = Q_{m+1} \cup \dots \cup Q_k \in \mathcal{F}$. Hence, $\textbf{Q}' = (Q_1, \dots , Q_{k-1}, Q_k \cup P^\textbf{c})$ is an $\mathcal{F}$-partition with $\textbf{B} \leq \textbf{Q}'$.
\qed
\begin{lem} If $\mathcal{F} = \mathcal{F}^\textbf{u} = \mathcal{F}^\textbf{s} $ and $\textbf{A} = (A_1, \dots A_k)$ is a classification problem then $ \textbf{A} \in \textbf{\textit{class}}_k(\mathcal{F})$ if and only if $ (A_i, A_j) \in \textbf{\textit{class}}_2(\mathcal{F})$ for all $ 1\leq i \neq j \leq k $.
\end{lem}

\proof 
The "if part" follows by Proposition 2.2. Suppose that $ (A_i, A_j) \in \textbf{\textit{class}}_2(\mathcal{F})$ for $ 1\leq i \neq j \leq k $. Now we proceed by induction over $|\textbf{A}| = k$. If k = 2 nothing is to prove. Let $\textbf{A} = (A_1, \dots ,A_{k+1})$  and suppose $ (A_1, \dots , A_k) \in \textbf{\textit{class}}_k(\mathcal{F})$. Then an $\mathcal{F}$-partition $\textbf{Q}' = (Q_1', \dots ,Q_k')$ with  $ (A_1, \dots , A_k) \leq \textbf{Q}'$ exists. Assume without loss of generality $A_i \subseteq Q_i'$ for $1 \leq i \leq k$. On the other side $Q_i'' \in \mathcal{F}^\textbf{dc}$ exist with $ A_i \subseteq Q_i''$ and $A_{k+1} \subseteq (Q_i'')^{\textbf{c}}$ for $1 \leq i \leq k$.
Consider $P = Q_1'' \cup \dots \cup Q_k''$. Then $A_i \subseteq P \in \mathcal{F}$ for $1 \leq i \leq k$ and $P^{\textbf{c}} = (Q_1'')^{\textbf{c}} \cap \dots \cap (Q_k'')^{\textbf{c}} \in \mathcal{F}$ with $A_{k+1} \subseteq P^{\textbf{c}}$. This shows $\textbf{Q} = (Q_1' \cap P, \dots , Q_k' \cap P, P^{\textbf{c}})$  is an $\mathcal{F}$-partition with $ \textbf{A} \leq \textbf{Q}$.
\qed

As indicated in the introduction we generalize the notion of a classification problem to conditional classification problems by fixing one component as condition. Consider $C \subseteq S$ and  a classification problem \textbf{A}. Then $(C, \textbf{A})$ is a \textit{conditional classification problem} if $C \cap \textbf{\textit{set}}(\textbf{A}) = \emptyset$, referring to $C$ as the \textit{problem condition}. $C$ could be finite, even empty. If $C^\textbf{c}$ is finite, then no conditional classification problems $(C, \textbf{A})$ exist.

\begin{defi} A conditional classification problem $(C,\textbf{A})$ is called $\mathcal{F}$-\textit{solvable} ($A\!\in\!\textbf{\textit{cclass}}_k(C, \mathcal{F})$) if and only if an $\mathcal{F}$-partition $\textbf{Q}= (Q_0, Q_1,\dots, Q_k)$ exists with $C \subseteq Q_0$ and $\textbf{A} \leq (Q_1,\dots, Q_k)$ where $k = |\textbf{A}|$.
\end{defi}
The following facts follow directly from the definition
\begin{prop} 
Let $\mathcal{F}$ and $k > 0$ be given.
\begin{enumerate}
\item
  $ C_1 \subseteq C_2 \subseteq S \ \Rightarrow\ \textbf{cclass}_k(C_2, \mathcal{F}) \subseteq \textbf{cclass}_k(C_1, \mathcal{F}) $.

\item
  $ C^\textbf{c} \in \textbf{fin}(S) \ \Rightarrow\ \textbf{cclass}_k(C, \mathcal{F}) = \emptyset $.

\item
 $ \emptyset \in \mathcal{F} \ \Rightarrow\ \textbf{class}_k(\mathcal{F}) \subseteq \textbf{cclass}_k(\emptyset, \mathcal{F})$.

\item
  $\mathcal{F} = \mathcal{F}^\textbf{u} \ \Rightarrow\ \textbf{class}_k(\mathcal{F}) = \textbf{cclass}_k(\emptyset, \mathcal{F})$.

\item
  $\mathcal{F}$ nontrivial and $ C \in \textbf{fin}(S) \ \Rightarrow\ \textbf{cclass}_k(C, \mathcal{F}) = \textbf{cclass}_k(\emptyset, \mathcal{F})$.
\end{enumerate}
\end{prop}

\begin{exa}
Consider $X = \{a,b\}$. Let $ \mathcal{L} = \mathcal{L}^\textbf{ltr} = \mathcal{L}^\textbf{-ltr}$ a nontrivial language family, which is closed under regular variation. If $A$ is a set with $A^\textbf{c}, A \notin \mathcal{L}$, then $(A^\textbf{c}, A) \notin \textbf{\textit{class}}_2(\mathcal{L})$ and by our assumption on $\mathcal{L}$ $(xA^\textbf{c}, xA) \notin \textbf{\textit{class}}_2(\mathcal{L})$ for $x = a,b$ (Lemma 5.4. in [4]). Clearly, $(aA^\textbf{c}, bA) \in \textbf{\textit{class}}_2(\mathcal{L})$, but $(aA \cup bA^\textbf{c}, aA^\textbf{c}, bA) \notin \textbf{\textit{class}}_3(\mathcal{L})$.
Hence $(aA^\textbf{c}, bA) \notin \textbf{\textit{cclass}}_2(aA \cup bA^\textbf{c}, \mathcal{L})$. 
\end{exa}

\section{Unsolvability Cores in Classification Problems}
As in the case of promise problems unsolvability of classification problems is closely related to cohesiveness.
\begin{defi} $A\subseteq S$ is $\mathcal{F}$-cohesive ($A \in \textbf{\textit{cohesive}}(\mathcal{F})$) if and only if $A$ is infinite and for all $ Q \in \mathcal{F}^\textbf{dc}$ either $ A \cap Q$ or $A \cap Q^{\textbf{c}}$ is finite (cf.[4] and [7]).
\end{defi}
\begin{rem}  It is interesting to compare our definition of cohesiveness with related classical definitions, as they are presented in [7]. Consider the families $\mathcal{L}_{\textbf{r.e.}}(X)^\textbf{cc}$, $\mathcal{L}_{\textbf{r.e.}}(X)$ and $\mathcal{L}_{\textbf{rec}}(X)$. Then $ L \in \textbf{\textit{cohesive}}(\mathcal{L}_{\textbf{r.e.}}(X)^\textbf{cc})$ if and only if $L$ is cohesive in the classical sense.  Moreover, $\textbf{\textit{cohesive}}(\mathcal{L}_{\textbf{rec}}(X)) = \textbf{\textit{cohesive}}(\mathcal{L}_{\textbf{r.e.}}(X))$, since a language $Q$ is recursive if and only if $Q$ and $Q^{\textbf{c}}$ are recursively enumerable. Furthermore the definition of recursively indecomposability coincides with the definition of $ \mathcal{L}_{\textbf{rec}}(X)$-cohesiveness. In [7] we also find the notion of indecomposability. $L$ is indecomposable if there exist no infinite sets $L_1, L_2  \in \mathcal{L}_{\textbf{r.e.}}(X)$ such that $L_1 \cap L_2 = \emptyset, L \subseteq L_1 \cup L_2, L \cap L_1$ is infinite and $L \cap L_2$ is infinite. Then we find the following results in [7]. If $ L \in \textbf{\textit{cohesive}}(\mathcal{L}_{\textbf{r.e.}}(X)^\textbf{cc})$ then it is indecomposable and any indecomposable $L$ is $\mathcal{L}_{\textbf{rec}}(X)$-cohesive. None of the converse implications hold.
\end{rem}
In [4] (Theorem 5.1.) it is proven, that for a promise problem $(A, B)$ and a nontrivial set family $\mathcal{F}$ $A \cup B \in \textbf{cohesive}(\mathcal{F})$ if and only if $ A, B \in \textbf{cohesive}(\mathcal{F}) $ and $ (A, B) \notin \textbf{class}_2(\mathcal{F})$.   
This result leads to a much stronger one. In the theory of complexity we find the notion of \textit{hard cores} inside those sets which can be computed with bounded ressources (time, space, e.t.c. [3]). Similarily, we can consider \textit{unsolvability cores} of classification problems which are not solvable.
\begin{defi} For $k > 1$ a classification problem $ \textbf{A}$ with $|\textbf{A}|= k$ is a \textit{k-core} of $\mathcal{F}$ ($\textbf{A} \in \textbf{\textit {core}}_k(\mathcal{F}))$ if and only if for all classification problems $\textbf{A}'$ with  $\textbf{A}' \leq \textbf{A}$ and $|\textbf{A}'|>1: \textbf{A}' \notin \textbf{\textit{class}}_{|\textbf{A}'|}(\mathcal{F})$.
\end{defi}
Clearly, any subproblem of a core is itself a core. This is especially true for subproblems, which are promise problems. This enables us to use the results about unsolvability cores for promise problems from [4].
\begin{lem} If $\mathcal{F} = \mathcal{F}^\textbf{u} $ and $\textbf{A} = (A_1, \dots, A_k) (k>1)$ is a classification problem then $ \textbf{A} \in \textbf{\textit{core}}_k(\mathcal{F})$ if and only if $ (A_i, A_j) \in \textbf{\textit{core}}_2(\mathcal{F})$ for all $ 1\leq i \neq j \leq k $.
\end{lem}
\proof Suppose $\textbf{A} \in \textbf{\textit {core}}_k(\mathcal{F})$, then by definition $(A_i, A_j) \leq \textbf{A}$ and therefore $ (A_i, A_j) \in \textbf{\textit {core}}_2(\mathcal{F}) $.  Conversely, suppose that $\textbf{A} \notin \textbf{\textit{core}}_k(\mathcal{F})$, i. e. $ \textbf{A}' = (A_1', \dots\ , A_m')$ exists with $ \textbf{A}' \leq \textbf{A}$, $m>1$ and $\textbf{A}' \in \textbf{\textit {class}}_{|\textbf{A}'|}(\mathcal{F})$. Since $\mathcal{F} = \mathcal{F}^\textbf{u} $ we know $ (A_1', A_2') \in \textbf{\textit{class}}_2(\mathcal{F})$. Moreover, $ A_1' \subseteq A_i$ and $ A_2' \subseteq A_j$ for some $ 1\leq i \neq j \leq k $. But then $(A_i, A_j) \notin \textbf{\textit{core}}_2(\mathcal{F})$.
\qed
Now we can characterize cores by cohesiveness. Using Theorem 5.1. and Theorem 6.7. of [4] we can prove
\begin{thm}If $\mathcal{F} = \mathcal{F}^\textbf{u} $ is nontrivial and \textbf{A} a classification problem with $|\textbf{A}|=k>1$ then $\textbf{A} \in \textbf{\textit{core}}_k(\mathcal{F})$ if and only if $ \textbf{\textit{set}}(\textbf{A}) \in \textbf{\textit{cohesive}}(\mathcal{F})$.
\end{thm}
\proof If $\textbf{A} = (A_1, \dots , A_k) \in \textbf{\textit {core}}_k(\mathcal{F})$, then $ (A_i, A_j) \in \textbf{\textit {core}}_2(\mathcal{F}) $ for all $ 1\leq i \neq j \leq k $. By Theorem 6.7. in [4] we know $A_1 \cup A_i \in \textbf{\textit{cohesive}}(\mathcal{F}) $ for all $ 2\leq i \leq k $. But then $A_1 \cup \dots\cup A_k = (A_1 \cup A_2) \cup \dots \cup (A_1 \cup A_k)$. Since $A_1 \subseteq (A_1 \cup A_i) \cap (A_1 \cup A_j)$ for all $ 2\leq i \neq j \leq k $ and $A_1$ is infinite, a simple induction proof shows $\textbf{\textit{set}}(\textbf{A}) \in \textbf{\textit{cohesive}}(\mathcal{F})$.

Conversely, if $ A_1 \cup \dots \cup A_k \in \textbf{\textit{cohesive}}(\mathcal{F})$ then for all $ 1\leq i \neq j \leq k$, $A_i \cup A_j \in \textbf{\textit{cohesive}}(\mathcal{F})$. Again by Theorem 6.7. of [4] $(A_i, A_j) \in \textbf{\textit{core}}_2(\mathcal{F})$ and therefore by Lemma 3.4. $\textbf{A} \in \textbf{\textit{core}}_k(\mathcal{F}).$
\qed
We can find to any classification problem $\textbf{A}$ with $|\textbf{A}| = 2$ and $ \textbf{A} \notin \textbf{\textit{class}}_2(\mathcal{F})$ a $\textbf{B} \leq \textbf{A}$ such that $\textbf{B} \in \textbf{\textit{core}}_2(\mathcal{F})$ if $\mathcal{F} = \mathcal{F}^\textbf{u} =\mathcal{F}^\textbf{s}$ is denumerable ([4]). But this is not true for classification problems $\textbf{A}$ with $|\textbf{A}| > 2$. To see this we prove the following theorem, where we use $S = X^*$ with $ X = \{a, b, c\}$. Define for $A \subseteq X^*$ the classification problem $\textbf{C}(A) = (A_{ab}, A_{bc}, A_{ca})$, where $A_{xy} = xA \cup yA ^\textbf{c}$ for $ x,y \in X$.
\begin{thm} Let $\mathcal{L}$ be a nontrivial language family with $ \mathcal{L} = \mathcal{L}^\textbf{u} = \mathcal{L}^\textbf{ltr} = \mathcal{L}^\textbf{-ltr}$, which is closed under regular variation. If $A \subseteq S$ with $A \notin \mathcal{L}$ or $A^{\textbf{c}} \notin \mathcal{L}$, then $\textbf{C}(A) \notin \textbf{\textit{class}}_3(\mathcal{L})$ and for all $\textbf{B} \leq \textbf{C}(A)$ with $|\textbf{B}| = 3 $ : $\textbf{B} \notin \textbf{\textit{core}}_3 (\mathcal{L})$.
\end{thm}
\proof (1) We know $(A^{\textbf{c}}, A) \notin \textbf{\textit{class}}_2(\mathcal{L})$ ([4]). But then by Lemma 5.4. of [4] $(xA^{\textbf{c}}, xA) \notin \textbf{\textit{class}}_2(\mathcal{L})$ for all $x \in X$. Now $(bA^{\textbf{c}}, bA) \leq (A_{ab}, A_{bc})$ , $(cA^{\textbf{c}}, cA) \leq (A_{bc}, A_{ca})$ and $(aA^{\textbf{c}}, aA) \leq (A_{ca}, A_{ab})$. This shows $ (A_{xy}, A_{xz}) \notin \textbf{\textit{class}}_2(\mathcal{L})$ for all $ x \neq y $ , $ z \neq y $ and $ x \neq z$.

(2) Suppose $\textbf{B} \leq \textbf{C}(A)$ exists with $ \textbf{B} \in \textbf{\textit {core}}_3(\mathcal{L}) $. Then by Theorem 3.5. $\textbf{\textit{set}}(\textbf{B}) \in \textbf{\textit{cohesive}}(\mathcal{L})$. Assume without loss of generality that $\textbf{B} = (B(a,b), B(b,c), B(c,a))$ and $B(x,y) \subseteq A_{xy}$ for $x,y \in X$ with $x \neq y$.. In the following let $B'(x, y) = B(x, y) \cap xX^*$ and $B''(x, y) = B(x, y) \cap (xX^*)^\textbf{c}$.

\textbf{Assertion} : $B'(x,y) \in \textbf{\textit{fin}}(X^*)$ for all $ x, y \in X$ with $ x \neq y $.\\
 Suppose to the contrary (without loss of generality) $B'(a,b) \notin \textbf{\textit{fin}}(X^*)$. But then $B'(b,c) \in \textbf{\textit{fin}}(X^*)$. Otherwise we obtain $(B'(a,b), B'(b, c)) \leq (aX^*, bX^*) \leq (aX^*, (aX^*)^\textbf{c})$. Since $\mathcal{L}_\textbf{reg} \subseteq \mathcal{L}$, $\textbf{B} \notin \textbf{\textit{core}}_3(\mathcal{L})$ - a contradiction. But now $B''(b,c)$ is infinite and $B''(b,c)\subseteq cX^* \subseteq (aX^*)^\textbf{c}$, hence both $\textbf{\textit{set}}(\textbf{B}) \cap aX^*$ and $\textbf{\textit{set}}(\textbf{B}) \cap (aX^*)^\textbf{c}$ are infinite - a contradiction to $\textbf{\textit{set}}(\textbf{B}) \in \textbf{\textit{cohesive}}(\mathcal{L})$.\\
Now consider $B''(a,b)$ and $B''(c,a)$. Then both sets are infinite and $(B''(a, b), B''(c,a)) \leq (bX^*, aX^*) \leq (bX^*, (bX^*)^\textbf{c})$ - a contradiction to $\textbf{B} \in \textbf{\textit{core}}_3(\mathcal{L})$. This completes the proof.
\qed

\begin{rem} The basic idea behind the proof of Theorem 3.6. is due to M. Ziegler ([1]). Note, that complexity classes and most of the known language families satisfy the conditions of Theorem 3.6.
\end{rem}

Using conditional unsolvability, we can derive an existence theorem for cores.
\begin{thm} Let $\mathcal{F}=\mathcal{F}^\textbf{u}=\mathcal{F}^\textbf{s}$ be denumerable and nontrivial. If $\textbf{A} = (A_1, \dots ,A_k)$ is a classification problem and $C \subseteq \textbf{\textit{set}}(\textbf{A})^\textbf{c}$ is $\mathcal{F}$-cohesive with $(C, A_i) \notin \textbf{\textit{class}}_2(\mathcal{F})$  for $ 1 \leq i \leq k$, then there exists $\textbf{B} \leq \textbf{A}$ with $|\textbf{B}| = k$ and $ \textbf{B} \in \textbf{\textit{core}}_k(\mathcal{F})$.
\end{thm}
\proof Since $(C, A_i) \notin \textbf{\textit{class}}_2(\mathcal{F})$, we can find $C_i \subseteq C$ and $B_i \subseteq A_i$ with $(C_i, B_i) \in \textbf{\textit{core}}_2(\mathcal{F})$ (Theorem 6.14. in [4]). By Theorem 3.5. $C_i \cup B_i \in \textbf{\textit{cohesive}}(\mathcal{F})$ and therefore $ B_i \in \textbf{\textit{cohesive}}(\mathcal{F})$. Now $(C, B_i) \notin \textbf{\textit{class}}_2(\mathcal{F})$ and $C  \in \textbf{\textit{cohesive}}(\mathcal{F})$. By Theorem 5.1. in [4] we know $C \cup B_i \in \textbf{\textit{cohesive}}(\mathcal{F})$. But then $C \cup B_1 \cup \dots \cup B_k = (C \cup B_1) \cup \dots \cup (C \cup B_k) \in \textbf{\textit{cohesive}}(\mathcal{F})$, since for all $1 \leq i\neq j \leq k$ $C$ is infinite and $C \subseteq (C \cup B_i) \cap (C \cup B_j)$. It follows $B_1 \cup \dots \cup B_k \in \textbf{\textit{cohesive}}(\mathcal{F})$ and we obtain $\textbf{B} = (B_1, \dots , B_k) \leq \textbf{A}$  and by Theorem 3.5. $ \textbf{B} \in \textbf{\textit{core}}_k(\mathcal{F})$.
\qed 

\begin{rem} Consider the situation of Theorem 3.6. Then $\textbf{\textit{set}}(\textbf{C}(A)) = XX^*$ and there is no room for an infinite condition $C$ to make the conditional classification problem $(C, \textbf{C}(A))$ $\mathcal{L}$-solvable.
\end{rem}

\section{Cores in Conditional Classification Problems}
Unsolvability of conditional classification problems can be related to cohesiveness, too.
\begin{defi} Let $C, A \subseteq S$. Then $A$ is $\mathcal{F}$\textit{-cohesive under condition} $C$ (in short: $A \in  \textbf{\textit{ccohesive}}(C, \mathcal{F}))$, if and only if $A$ is infinite and for all $Q \in \mathcal{F}^{\textbf{dc}}$ with $ Q \subseteq C$ either $ A \cap Q$ or $ A \cap Q^c $ is finite.
\end{defi}
Clearly, if $ C_1 \subseteq  C_2 \subseteq S$, then \textbf{\textit{ccohesive}}$(C_2, \mathcal{F}) \subseteq$ \textbf{\textit{ccohesive}}$(C_1, \mathcal{F})$. 
Especially, we get $\textbf{\textit{ccohesive}}(S, \mathcal{F})=\textbf{\textit{cohesive}}(\mathcal{F})$ and therefore $\textbf{\textit{cohesive}}( \mathcal{F}) \subseteq \textbf{\textit{ccohesive}}(C, \mathcal{F})$ for all $C \subseteq S$. Rewriting the definition, we also find $\textbf{\textit{ccohesive}}(C, \mathcal{F})) = \textbf{\textit{cohesive}}(\mathcal{F}(C)^\textbf{cc})$ where $\mathcal{F}(C) = \{Q |\ Q \subseteq C$ and $ Q \in \mathcal{F}\}$.  Analogously, we define conditional cores by
\begin{defi} Let $C \subseteq S$ and $\textbf{A}$ a classification problem. Then $\textbf{A}$ is a $C$-$\textit{conditional}\ \textit{core}$ of $\mathcal{F}$ 
($\textbf{A} \in \textbf{\textit {ccore}}_{|\textbf{A}|}(C, \mathcal{F}))$ if and only if for all $\textbf{A}' \leq \textbf{A}$ with $|\textbf{A}'|>0: \textbf{A}' \notin \textbf{\textit{cclass}}_{|\textbf{A}'|}(C, \mathcal{F})$.
\end{defi}
In contrast to the definition of $\textbf{\textit{core}}(\mathcal{F})$ subproblems $\textbf{A}'$ with $|\textbf{A}'| = 1$ are considered, too.
Note, that $(C, \textbf{A}')$ is a conditional-classification problem, if $\textbf{A}' \leq \textbf{A}$. Moreover, if $\textbf{A} \in \textbf{\textit{ccore}}_{|\textbf{A}|}(C, \mathcal{F})$, then $\textbf{A}' \in \textbf{\textit{ccore}}_{|\textbf{A}'|}(C, \mathcal{F})$. The following lemma characterizes $A \in \textbf{\textit{ccore}}_1(C, \mathcal{F})$ by conditional cohesiveness.
\begin{lem} Let $\mathcal{F}$ be nontrivial and $C, A \subseteq S$ with $A$ infinite and $A \cap C = \emptyset$. Then the following statements are equivalent
\begin {enumerate}[label=(\roman*)]
\item
$ A \in \textbf{\textit {ccore}}_1(C, \mathcal{F})$
\item
$A \notin \textbf{\textit{cclass}}_1(C, \mathcal{F})$ and $ A \in \textbf{\textit{ccohesive}}(C^c, \mathcal{F})$.
\end{enumerate} 
\end{lem}
\proof 
(i) $\Rightarrow$ (ii): Suppose $ A \in \textbf{\textit {ccore}}_1(C, \mathcal{F})$. Then $A \notin \textbf{\textit{cclass}}_1(C, \mathcal{F})$. Assume to the contrary that $ A \notin \textbf{\textit{ccohesive}}(C^{\textbf{c}}, \mathcal{F})$. 
Then $Q \in \mathcal{F}^{\textbf{dc}}$ exists with $ Q \subseteq C^{\textbf{c}}, A \cap Q \notin \textbf{\textit{fin}}(S)$ and $ A\cap Q^{\textbf{c}} \notin \textbf{\textit{fin}}(S)$. 
Let $B = A \cap Q$. Then $ B \subseteq Q$, but $Q \subseteq C^{\textbf{c}}$, hence $C \subseteq Q^{\textbf{c}}$. Moreover, $ Q, Q^{\textbf{c}} \in \mathcal{F}$, i.e. $B \in \textbf{\textit{cclass}}_1(C, \mathcal{F})$.

(ii) $\Rightarrow$ (i): Suppose that $A \notin \textbf{\textit{cclass}}_1(C, \mathcal{F})$ and $ A \in \textbf{\textit{ccohesive}}(C^{\textbf{c}}, \mathcal{F})$. Assume to the contrary that an infinite set $B \subseteq A$ exists, such that $B \subseteq Q^{\textbf{c}}$ and $C \subseteq Q$ for some $Q \in \mathcal{F}^{\textbf{dc}}$. Then  $Q^{\textbf{c}} \subseteq C^{\textbf{c}}$. Since $B \cap Q^{\textbf{c}} \notin \textbf{\textit{fin}}(S)$, $A \cap Q^\textbf{c} \notin \textbf{\textit{fin}}(S)$, too.  Hence $A \cap Q \in \textbf{\textit{fin}}(S)$, because $ A \in \textbf{\textit{ccohesive}}(C^{\textbf{c}}, \mathcal{F})$. Consider $Q'=Q^{\textbf{c}} \cup (A \cap Q)$. Since $\mathcal{F}$ is nontrivial, $Q' \in \mathcal{F}$. Note that $A=(A \cap Q) \cup (A \cap Q^{\textbf{c}}) \subseteq Q^{\textbf{c}} \cup (A \cap Q)= Q'$. On the other side, $Q^{\textbf{c}} \subseteq C^c$ and $A \cap Q \subseteq A \subseteq C^{\textbf{c}}$, i.e. $Q' \subseteq C^{\textbf{c}}$. Hence $C \subseteq Q'^{\textbf{c}}$. This shows that $A \notin \textbf{\textit{cclass}}_1(C, \mathcal{F})$ - a contradiction.
\qed
\begin{thm} Let $ \mathcal{F}$ be nontrivial with $ \mathcal{F}=\mathcal{F}^\textbf{u}$ and $(C, \textbf{A})$ a conditional k-classification problem. If $\textbf{A}=(A_1, \dots, A_k)$ then the following statements are equivalent
\begin{enumerate}[label=(\roman*)]
\item
$ \textbf{A} \in \textbf{\textit {ccore}}_k(C, \mathcal{F})$
\item
$A_i \notin \textbf{\textit{cclass}}_1(C, \mathcal{F})$ and $ A_i \in \textbf{\textit{ccohesive}}(C^{\textbf{c}}, \mathcal{F})$ for all $1 \leq i \leq k$.
\end{enumerate} 
\end{thm}
\proof\hfill

(i) $\Rightarrow$ (ii): Suppose that  $\textbf{A} \in \textbf{\textit {ccore}}_k(C, \mathcal{F})$. Then for all $1\leq i \leq k: (C, A_i) \in \textbf{\textit{ccore}}_1(C, \mathcal{F})$, since $A_i \leq \textbf{A}$. Applying Lemma 4.3. we get the result.

(ii) $\Rightarrow$ (i): Let the $A_i$ be given according to the assumption. Assume to the contrary that $\textbf{B} \leq \textbf{A}$ exists with $\textbf{B}=(B_1, \dots, B_m) \in \textbf{\textit{cclass}}_m(C, \mathcal{F})$. Then an injective $ \sigma: [m] \rightarrow [k]$ exists with $B_i \subseteq A_{\sigma (i)}$ for $1 \le i \leq k$. Since $ \mathcal{F} =\mathcal{F}^\textbf{u}, B_i \in  \textbf{\textit{cclass}}_1(C, \mathcal{F})$. But $ A_{\sigma (i)} \in \textbf{\textit{core}}_1(C, \mathcal{F})$ and $ B_i \subseteq  A_{\sigma (i)}$. This is a contradiction.
\qed 
Now, we are able to assert the existence of conditional cores in the case that both $C$ and $C^{\textbf{c}}$ are infinite. 
Observe that under this assumption $A \in \textbf{\textit{cclass}}_1(C, \mathcal{F})$ if and only if $(C, A)$ considered as a promise problem is solvable for $ \mathcal{F}$, i.e. $(C,A) \in \textbf{\textit{class}}_1(\mathcal{F})$. 
\begin{lem}
Let $ \mathcal{F}$ be denumerable and nontrivial with $\mathcal{F}=\mathcal{F}^\textbf{u}=\mathcal{F}^\textbf{s}$. If $A \notin \textbf{\textit{fin}}(S)$, $C \notin \textbf{\textit{fin}}(S)^\textbf{cc}$, $A \cap C = \emptyset$ and $A \notin \textbf{\textit{cclass}}_1(C, \mathcal{F})$, then $B \subseteq A$ exists with $B \in \textbf{\textit{ccore}}_1(C, \mathcal{F})$.
\end{lem}
\proof
If $A \notin \textbf{\textit{cclass}}_1(C, \mathcal{F})$, i.e. $ (C, A) \notin \textbf{\textit{class}}_1(\mathcal{F})$. By cor.5.16. in[4] we can find $B \subseteq A$ such that for all infinite $B' \subseteq B$ $(C, B') \notin \textbf{\textit{class}}_2(\mathcal{F})$, i.e. $B \in \textbf{\textit{ccore}}_1(C, \mathcal{F})$.   
\qed
Using this lemma in connection with Theorem 4.4. we get
\begin{lem}
Let $\mathcal{F}$ be denumerable and nontrivial with  $\mathcal{F}=\mathcal{F}^\textbf{u}=\mathcal{F}^\textbf{s}$ and $(C,\textbf{A})$ a conditional classification problem where $C$ and $C^{\textbf{c}}$ are infinite. If $\textbf{A}=(A_1, \dots, A_k)$ with $A_i \notin \textbf{\textit{cclass}}_1(C,\mathcal{F})$ for $ 1\leq i \leq k$ then a $ \textbf{B} \leq \textbf{A}$ exists with $|\textbf{B}|=k$ and $ \textbf{B} \in \textbf{\textit{ccore}}_k(C,\mathcal{F})$.
\end{lem}
\proof
By Lemma 4.5. we find for each $1\leq i \leq k$ $B_i \in \textbf{\textit{ccore}}_1(C,\mathcal{F})$ and $B_i \subseteq A_i$. Let $\textbf{B}=(B_1, \dots, B_k)$. Then $\textbf{B} \leq \textbf{A}$ and $|\textbf{B}|=k$. By Theorem 4.4. $\textbf{B} \in \textbf{\textit{ccore}}_k(C,\mathcal{F})$.
\qed
\section{Conditional Cores and Hard Cores}
For WP-recursive language families we can prove a much stronger result. This depends on the relation between $A \in \textbf{\textit{ccore}}_1(C,\mathcal{F})$ and proper hard cores introduced by N. Lynch [6] for complexity classes and in a very general form by R. Book- D.-Z. Du [3]. 
\begin{defi}
$B$ is a $\mathcal{F}$-\textit{hardcore} of $A$ if and only if $B$ is infinite and for all $C \in \mathcal{F}(A)$: $ B \cap C \in \textbf{\textit{fin}}(S)$. If additionally $B \subseteq A$ then $B$ is a \textit{proper} $\mathcal{F}$-hardcore of $A$. (Remind $ \mathcal{F}(A)= \{ Q \subseteq A\ |\ Q \in \mathcal{F}\}$ for $\mathcal{F}$ and $A$.)
\end{defi} 
Note, that for $A' \subseteq A$ with $A'$ infinite every $\mathcal{F}$-hardcore of $A$ is a $\mathcal{F}$-hardcore of $A'$. Rephrasing Lemma 7.2. of [4] we get the following
\begin{lem}
If $ \mathcal{F}$ is nontrivial with $\mathcal{F}=\mathcal{F}^\textbf{co}$ and $(C, A)$ a conditional classification problem then $A$ is a proper $\mathcal{F}$-hardcore of $C^{\textbf{c}}$ if and only if $A \in \textbf{\textit{ccore}}_1(C,\mathcal{F})$.
\end{lem}
Now we can use a construction for proper hard cores from [3] in a modified form.
\begin{thm}
If $ \mathcal{L}$ is a nontrivial and WP-recursive language family with $\mathcal{L}= 
\mathcal{L}^\textbf{b}$ and $(C, A)$ a conditional classification problem with $A \notin \textbf{\textit{cclass}}_1(C,\mathcal{L})$ and $ C, A$ are recursive then a recursive $B \subseteq A$ exists with $B \in \textbf{\textit{ccore}}_1(C,\mathcal{L})$.
\end{thm}

\proof
Consider an enumeration $\textbf{e}$ of $\mathcal{L}$ such that $\textbf{\textit{word}}_\textbf{e} \in \textbf{\textit{rec}}_2$.
Furthermore, let $\delta_C, \delta_A \in \textbf{\textit{rec}}_1$. Now define for all $n \geq 0$ $B(n), \text{cancel}(n)$ and $ \text{card}(n)$ by the following algorithm:\\
\begin{algorithmic}
\IF{$\textbf{\textit{lex}}(0) \in C$}
\STATE $\text{cancel}(0):=0$
\ENDIF
\IF{$\textbf{\textit{lex}}(0) \in A$ and $\textbf{\textit{lex}} \notin  \textbf{e}(0)$}
\STATE $B(0):=0$; $\text{card}(0):=1$
\ENDIF
\STATE $n:= 1$;
\WHILE {$n \neq 0$}
\STATE $B(n):= B(n-1)$; $ \text{cancel}(n):= \text{cancel}(n-1)$; $\text{card}(n):= \text{card}(n-1)$;
\IF{$\textbf{\textit{lex}}(n) \in C$}
\STATE $\text{cancel}(n):=\text{cancel}(n) \cup \{i | 0 \leq i \leq \text{card}(n)$ and $ \textbf{\textit{lex}}(n) \in \textbf{e}(i)\}$
\ENDIF
\IF{$\textbf{\textit{lex}}(n) \in A$ and $\forall \ 0 \leq i \leq \text{card}(n): (i \notin \text{cancel}(n) \Rightarrow \textbf{\textit{lex}}(n) \notin \textbf{e}(i))$ }
\STATE $B(n):=B(n) \cup \textbf{\textit{lex}}(n); \text{card}(n):= \text{card}(n)+1$ 
\ENDIF;
\STATE $n:=n+1$
\ENDWHILE 
\end{algorithmic}

(For $A=C^{\textbf{c}}$ we get the construction of [3]). \\

Now, let $ B= \bigcup \nolimits _{i=0}^\infty B(n)$ and $\text{cancel}= \bigcup \nolimits  _{i=0}^\infty \text{cancel}(i)$. Assume for the moment that $B$ is infinite. $B$ is recursive and $B \subseteq A$, since all basic functions are recursive, $\text{cancel}(n)$ is finite for all $n$ and the elements of $B$ are added in increasing order with respect to $\textbf{\textit{lex}}$. Moreover, $\text{lim}_{n \rightarrow \infty} \text{card}(n)= \infty$. Hence $\{k | \textbf{e}(k) \cap C \neq \emptyset\} = \text{cancel}$ and we get $\textbf{e}(i) \subseteq C^{\textbf{c}}$ 
and by construction $ \textbf{e}(i) \cap B \in \textbf{\textit{fin}}(X^*)$ for $ i \notin \text{cancel}$ (cf. [3]). In conclusion, $B$ is a proper $\mathcal{L}$-hardcore of $C^{\textbf{c}}$ and by Lemma 4.9. $B  \in \textbf{\textit{ccore}}_1(C,\mathcal{L})$.
It remains to show the 

\textbf{Assertion:} $B \notin \textbf{\textit{fin}}(X^*)$.\\
Suppose to the contrary, that $B$ is finite. Then $M$ exists with $card(n)=M$ for almost all $n$. Moreover, for every $i \in [M+1]_0$ with $\textbf{e}(i) \cap C \neq \emptyset$ there must exist $K(i)$ with $ i \in \text{cancel}(K(i))$. Let $K= \textbf{\textit{max}}\{K(i) | i \in [M+1]_0$ with $ \textbf{e}(i) \cap C \neq \emptyset\}$. Then we know that for all  $i \in [M+1]_0$ with $i \notin \text{cancel}(K(i)): \textbf{e}(i) \subseteq C^{\textbf{c}}$. Choose $ N \geq K$
sufficiently large such that additionally card$(n)=M$ for every $ n \geq N$.
Consider $ \textbf{\textit{lex}}(n) \in A$ with $n \geq N$. Since $ \textbf{\textit{lex}}(n) \notin B$, $i \in [M+1]_0$ exists with $ \textbf{\textit{lex}}(n) \in \textbf{e}(i)$. 
This shows $A \subseteq \{ \textbf{\textit{lex}}(k) | k < N$ and 
$\textbf{\textit{lex}}(k)  \in A\} \cup 
\bigcup \nolimits _{i=0, i \notin \text{cancel}}^M
\textbf{e}(i)=Q \subseteq C^\textbf{c}$ and therefore $C \subseteq Q^\textbf{c}$.
Since $\mathcal{L}$ is nontrivial and $\mathcal{L}=\mathcal{L}^\textbf{u}$, we know $Q \in \mathcal{L}$. Moreover, $\mathcal{L}= \mathcal{L}^\textbf{co}$ implies $Q^{\textbf{c}} \in \mathcal{L}$, hence $A \notin \textbf{\textit{cclass}}_1(C,\mathcal{L})$ - a contradiction. 
\qed
Now we can derive a stronger result than Lemma 4.6.:
\begin{thm}
Let $\mathcal{L}$ be a nontrivial and WP-recursive language family with $\mathcal{L}= \mathcal{L^\textbf{b}}$ and $(C, \textbf{A})$ a conditional k-classification problem. If $C$ is recursive and $\textbf{A}=(A_1, \dots, A_k)$ such that $A_i \in \textbf{\textit{cclass}}_1(C, \mathcal{L})$ and $A_i$ is recursive for $1 \leq i \leq k$ then $\textbf{B}=(B_1, \dots, B_k)$ exists with $\textbf{B} \leq \textbf{A}, \textbf{B} \in \textbf{\textit{ccore}}_k(C, \mathcal{L})$ and $B_i$ is recursive for $1 \leq i \leq k$.
\end{thm}
\proof
By Theorem 5.3. we find for each $1 \leq i \leq k$  $B_i \in \textbf{\textit{cclass}}_1(C, \mathcal{L})$ with $B_i \subseteq A_i$ and $B_i$ is  recursive. Let $\textbf{B}=(B_1, \dots, B_k)$. Then $\textbf{B} \leq \textbf{A}$ and by Theorem 4.4. $\textbf{B} \in \textbf{\textit{ccore}}_k(C, \mathcal{L})$. 
\qed
\begin{rem} The $B_i$'s constructed in Theorem 5.4. are all infinite. By the Dekker-Myhill theorem (\S 12.3 Theorem VI in [7]), we can find in every $B_i$ a $\mathcal{L}$-cohesive $B_i'$, but we cannot show, that $B_i'$ is recursive under the conditions of Theorem 5.4. The best result to our knowledge  is the result of Friedberg (\S 12.4 Theorem XI in [7]). The construction (due to Yates) in the proof given in [7] can be easily modified in such a way, that to any infinite, recursive $A$ a $\mathcal{L}_\textbf{r.e.}(X)$-cohesive subset $B$ with $B^\textbf{c} \in \mathcal{L}_\textbf{r.e.}(X)$ can be found. Since any WP-recursive language family $\mathcal{L}$ is a subfamily of $\mathcal{L}_\textbf{r.e.}(X)$ this $B$ is $\mathcal{L}$-cohesive, too.
\end{rem}

\section*{Concluding Remarks}
This paper continues our research about unsolvability cores in promise problems ([4]) generalizing the results to classification problems. Our approach is very general, though the applications in this paper deal mainly with language families and complexity classes. The main open problem in our approach is to construct cohesive sets with "nice" properties.

\section*{Acknowledgment}
  \noindent We acknowledge seminal discussions with Martin Ziegler.

\section*{references}
\begin{enumerate}
\item[[1]]
\textsc{K. Ambos-Spies, U. Brandt, M. Ziegler}:
  ``Real Benefit of Promises and Advice'', pp. 1-11, CiE 2013
  \emph{}
\item[[2]]
\textsc{J.L.Balcazar, J.Diaz, J.Gabarro}:
  `` Structural Complexity I" 
  \emph{EATCS Monographs on Theoretical Computer Sciences}, Springer Verlag (1988)
\item[[3]] 
  \textsc{R.V.Book, Ding-Zhu Du}:
  ``The Existence and Density of Generalized Complexity Cores'', pp.718--730 in
  \emph{JACM} vol.\textbf{34:3} (July 1987).
\item[[4]] 
  \textsc{U.Brandt, H.K.-G.Walter}:
  ``Cohesiveness in Promise Problems'',
  \emph{RAIRO - Theoretical Informatics and Applications} vol. \textbf{47:4} (November 2013).
\item[[5]]
  \textsc{S.Even, A.L.Selman and Y.Yacobi}:
  ``The Complexity of Promise Problems with Applications to Public-Key Cryptography'', pp.159--173 in
  \emph{Information and Control} vol.\textbf{61} (1984).
\item[[6]]
\textsc{N.Lynch}:
  `` On Reducibility to Complex or Sparse sets" 
  \emph{JACM},  vol.\textbf{3}(July 1975), pp.341-345
\item[[7]]
  \textsc{Hartley Rogers jun.}:``\emph{Theory of Recursive Functions and Effective Computability}'',
  MacGraw-Hill Book Company (1967).  
\end{enumerate}

\end{document}